# Measuring Book Impact Based on the Multi-granularity Online Review Mining

Qingqing Zhou[1, 3], Chengzhi Zhang [1, 2, 3, *], Star X. Zhao[4], Bikun Chen[1]

1. Department of Information Management, Nanjing University of Science and Technology, Nanjing, 210094, China
2. Jiangsu Key Laboratory of Data Engineering and Knowledge Service (Nanjing University), Nanjing, 210094, China
3. Alibaba Research Center for Complex Sciences, Hangzhou Normal University, Hangzhou, 311121, China
4. Department of Information Science, Business School, East China Normal University, Shanghai, 200241, China

**Abstract.** As with articles and journals, the customary methods for measuring books' academic impact mainly involve citations, which is easy but limited to interrogating traditional citation databases and scholarly book reviews. Researchers have attempted to use other metrics, such as Google Books, libcitation, and publisher prestige. However, these approaches lack content-level information and cannot determine the citation intentions of users. Meanwhile, the abundant online review resources concerning academic books can be used to mine deeper information and content utilizing altmetric perspectives. In this study, we measure the impacts of academic books by multi-granularity mining online reviews, and we identify factors that affect a book's impact. First, online reviews of a sample of academic books on Amazon.cn are crawled and processed. Then, multi-granularity review mining is conducted to identify review sentiment polarities and aspects' sentiment values. Lastly, the numbers of positive reviews and negative reviews, aspect sentiment values, star values, and information regarding helpfulness are integrated via the entropy method, and lead to the calculation of the final book impact scores. The results of a correlation analysis of book impact scores obtained via our method versus traditional book citations show that, although there are substantial differences between subject areas, online book reviews tend to reflect the academic impact. Thus, we infer that online reviews represent a promising source for mining book impact within the altmetric perspective and at the multi-granularity content level. Moreover, our proposed method might also be a means by which to measure other books besides academic publications.

**Keywords:** Online book reviews; Sentiment analysis; Book citation; Information content; Altmetrics

## Introduction

Impact measures pertaining to academic publications usually focus on research articles. However, monographs are also an important form of academic output. Book impact measures, traditionally, are similar to those used for articles, journals, or other research, and mainly take citations into consideration. For example, Su (2009) analyzed the impact of social sciences and humanities books using the Chinese Social Sciences Citation Index. However, with the development of Web 2.0, researchers have increasingly applied alternative metrics to assess the qualities of academic publications. Kousha et al. (2011) examined whether online citations from Google Books and Google Scholar could provide alternative sources of citation evidence. However, prior research based on citation metrics has often ignored the content of the information, which encompasses the intention and motivation of its users. Thus, deeper intention cannot be mined through such methods. For example, sometimes a negative citing is counted as a positive citation. In the present case, book impact assessments based on citations may not be accurate enough.

In a bid to try to assess the academic impact of books more comprehensively, some researchers have endeavored to combine citations metrics with scholarly reviews about the books under consideration. For example, Nicolaisen (2002) proposed a bibliometric technique for determining the scholarliness of scholarly book reviews. Zuccala et al. (2014) employed a machine-learning approach to qualitatively code scholarly book reviews as quality indicators to assess a book's impact. Although such analysis indicates that it is useful to combine scholarly book reviews with citations, the large-scale peer review

---

exercises required are expensive and time consuming. Also, existing research has neglected to incorporate sentiment information from the scholarly book reviews. Instead, online book reviews are abundant and widely available from e-commerce and social network websites, such as Amazon, LibraryThing, and Douban[1]. Compared with scholarly reviews, online reviews are plentiful and feature preferences expressed by users, which comprehensively and instantly reflect a book's impact. Hence, it may be worthwhile utilizing online reviews as an alternative impact measure.

In our study of measuring book impact in the online context, we propose a method based on multi-granularity mining of books' online reviews. This method aims to assess the impacts of academic books by mining online books reviews, and determining the most influential factors. We use online reviews of academic books as our dataset, and conduct multi-granularity mining on books reviews, in which macro-level review mining is used to identify the sentiment polarities of reviews and micro-level review mining is applied to calculate the sentiment values of aspects. Then, we apply the entropy method to integrate the values measured by review mining and compute the final book impact scores. In order to prove the validity of our method, correlation analysis between book impact scores obtained using our method and using traditional book citations is undertaken with the resultant significant correlations suggesting that online reviews can be used to measure book impact from an altmetric perspective.

**Related works**
In this section, we describe three categories of related works: (1) traditional impact assessments, (2) alternative impact assessments, and (3) online review mining.

**Traditional impact assessments**
Traditional impact assessments are mainly based on citations. Garfield (1972) used citation analysis as a tool in journal assessment, and found that journals can be ranked by frequency and impact of citations. Stremersch et al. (2007) demonstrated that citations were drivers of article impact by contrasting, synthesizing, and simultaneously testing three scientometric perspectives on the impact of article and author characteristics on article citations. Few prior studies, though, have paid much attention to book impact assessments.

Torres-Salinas et al. (2012) analyzed different impact indicators referred to by scientific publishers and included in the Book Citation Index for the social sciences and humanities fields during 2006–2011. Also, with the development of Thomson Reuters' Web of Science, Elsevier's Scopus, and the aforementioned Book Citation Index (again maintained by Thomson Reuters), many researchers have begun to use these and similar citation indexing services to assess book impact. Bar-Ilan (2010) examined three citation databases (Google Scholar, Scopus, and Web of Science) through citations of the book *Introduction to Informetrics* published by Leo Egghe and Ronald Rousseau (1990), so as to identify similarities and differences between the results obtained through them. Studies such as these show that citations can be a valuable measurement for evaluating books, as well as monographs. However, it represents just the viewpoint of traditional and offline impact.

**Alternative impact assessments**
Traditional, citation-based bibliometric methods are proving increasingly inadequate in the age of Web 2.0, and many researchers are seeking alternative metrics with which to assess various qualities of academic publications. Using the Web for research assessment, Kousha et al. (2010) introduced a new, combined, Integrated Online Impact indicator, and concluded that it can be used to help monitor research performance. Torres-Salinas et al. (2012) analyzed the different impact indicators referred to by book publishers in the Book Citation Index, while Gorraiz et al. (2013) introduced the Book Citation Index in detail, and Gorraiz et al. (2014) found that book reviews can be considered a suitable selection criterion for such a citation index.

Bornmann (2014) set out to ascertain whether altmetric data could validly be used for the measurement of societal impact with a comprehensive dataset from disparate sources. Shema et al. (2014) examined blog posts aggregated at ResearchBlogging.org, and, based on their results, suggested that blog citations could be used as an altmetric source. Torres-Salinas et al. (2014) used the Book Citation Index to analyze factors that determine the citation characteristics of books. Zuccala et al. (2015) assessed the value of reader ratings in Goodreads for measuring the wider impact of scholarly books published in the field of history. Their findings showed that, Goodreads, as a unique altmetric data source, could allow scholarly authors from the social sciences and humanities disciplines to measure the wider impact of their books. Haustein et al. (2015) discussed social media metrics in scholarly communication. Kousha and Thelwall (2015) assessed whether academic reviews in *Choice* (published by the U.S. Association of College and Research Libraries) could be systematically used for indicators of scholarly impact, uptake, or educational value for scholarly books. Their findings showed that metrics derived from *Choice* academic book reviews could be used as indicators of different aspects for books, but more evidences were needed before they could be used as proxies for peer judgments of individual books.

---
[1] http://www.douban.com (in Chinese).

These previous studies of alternative impact assessments reveal that, beyond the traditional methods like citations, more interesting perspectives of book impact measurement might be abstracted by using online or altmetric information.

**Online review mining**
In order to assess book impact via online reviews from e-commerce or social network websites, a data mining method is applied to process these reviews. As a burgeoning research topic, review mining has already attracted a comprehensive set of theories and technologies. For example, Shi and Chang (2006) extracted product feature-orientation (sentiment) pairs from online product reviews, and Ding et al. (2008) determined the semantic orientations (positive, negative, or neutral) of opinions expressed on product features in reviews using a holistic lexicon-based approach. Chaovalit and Zhou (2005) compared supervised and unsupervised classification approaches to mine movie reviews, while Zhuang et al. (2006) integrated WordNet, statistical analysis, and movie knowledge to determine whether opinions were positive or negative. Specifically regarding books, Chevalier and Mayzlin (2006) examined the effect of consumer reviews on the relative sales of books on Amazon.com and BarnesandNoble.com, mentioning nothing about book impact. Conversely, Kousha and Thelwall (2016) assessed whether a number of simple metrics derived from Amazon.com reviews of academic books could give evidence about their respective impact.

Previous research on the academic impact of books has generally considered traditional citation databases, scholarly book reviews, and so on. Recently, Thelwall and Kousha (2015) did describe web indicators for the impact of books, such as Google Books, Libcitations, book reviews, online book reviews, and book review sentiments, but, as suggested above, most of the current research continues to rely on the data of the scientific literatures, which commonly either lacks content information or neglects reviews from e-commerce or social network websites. Some prior studies, such as Shaw (1991) and Kousha and Thelwall (2016), have paid attention to book review sentiments, but they generally neglect the fine-grained sentiments about aspects of books.

In the present study, we aim to measure the impact of academic books using multi-granularity mining on online book reviews of Amazon.cn (Amazon China). In contrast to Kousha and Thelwall's (2016) study, we conduct micro-level sentiment analysis to identify the most influential factors, and focus on the context and content of online book reviews. We expect that the method presented herein could lead to new insights into content-level evaluation of book impact.

**Research questions**
This study aims to introduce a method for measuring academic book impact based on online content and sentiment analysis. In the empirical part of the study, we also try to evaluate whether online reviews are useful for the impact assessment of academic books. Additional research questions are addressed as follows:
- Are online book reviews a sufficient academic book impact measure?
- Which category of aspects most affects book impact: content-related aspects, publisher-related aspects, or operator-related aspects? For example, content, price, and packaging belong to these three categories, respectively. If a book is published well or sold on a good e-commerce platform, it may be accessed and read by more people. Hence, publisher-related aspects and operator-related aspects need to be considered as well as content-related aspects.
- Do disciplinary differences affect the answers to the questions above?

**Methodology**
**Data collection**
Academic books were used as the study's research samples, and, along with their citations, were selected using Su's (2011) "A report on the academic impact of Chinese books in the humanities and social sciences," which covers 20 academic disciplines of Chinese books from the humanities and social sciences fields. References to each book in the report include its academic discipline, title, author, publication year, and citation, with the latter collected from the Chinese Social Sciences Citation Index[2].

---
[2] http://cssci.nju.edu.cn/

For the present study, 544 economics books, 216 management books, 190 library and information science (LIS) books, 137 psychology books, and 428 literature books were chosen.

As noted above, most traditional reviews in this context are scholarly book reviews. Such appraisals are prefaced by the book's title, author, publisher, page extent, and price, after which the main content of the review is presented, including a content evaluation of the book and an academic impact evaluation; lastly, references and information about the review's author are presented. By contrast, in this study, the sample book reviews were taken from Amazon China (Amazon.cn), where they often differ in length and focus on diverse aspects of the same book. Fig. 1 shows a sample review of an edition of Karl Marx's *Das Kapital* (*Capital* in English), including the book's star rating (average customer rating using Amazon's 1–5 star scale), the review's contents, and a measure of its helpfulness. Star ratings reflects the overall assessment of a book by Amazon users (hereafter, "review holders"); a review's content is the main body of a book's review, reflecting users' intentions, sentiments, and assessments of the book; a review's helpfulness is judged by other users (hereafter, "review evaluators") in terms of whether or not it proved useful for prospective readers/purchasers (Yin et al. 2014). Other Amazon users—review evaluators—can assess a book review's helpfulness by clicking on the "Yes" or "No" voting buttons, as applicable. Thereby, review helpfulness can be used to evaluate star ratings and review contents, potentially reducing the effects of fake and low-quality reviews.

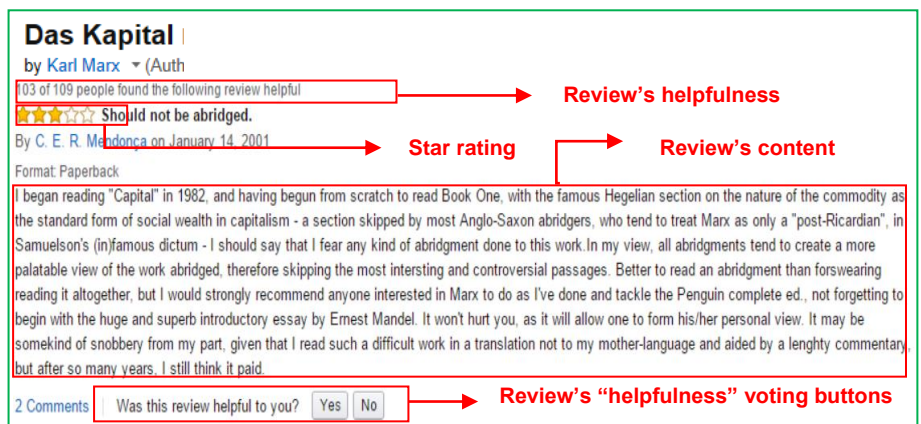

**Fig. 1.** An example of a book review of *Das Kapital* from Amazon.com

In order to use online reviews to evaluate the academic impact of the study's sample books, we crawled the reviews of each candidate book listed on Amazon.cn in October 2014. Each review had to contain three parts—"star rating," "review contents," and "review helpfulness"—and books with more than 10 reviews were extracted. Ultimately, 242 books, comprising 40 economics titles, 44 management titles, 30 psychology titles, 30 LIS titles, and 98 literature titles, were selected, the basic descriptive statistics for which are shown in Table 1.

**Table 1.** Descriptive statistics of the study's sample (after filtering via review frequency)

| Academic discipline | Initial number of books | Final number of books | Total number of book reviews | Average number of book citations |
|---|---|---|---|---|
| Economics | 544 | 40 | 3,002 | 213.50 |
| Management | 216 | 44 | 3,179 | 161.18 |
| LIS | 190 | 30 | 4,731 | 68.27 |
| Psychology | 137 | 30 | 2,302 | 21.90 |
| Literature | 428 | 98 | 8,041 | 165.71 |

**Table 2.** Sample books after information integration

| Academic discipline | Book title [a] | Sample review content | Star rating (1–5) | Review helpfulness | Citation |
|---|---|---|---|---|---|
| Economics | *Macroeconomics* | Such a translation level is offensive to authors and readers. | 1 | 35/40 | 143 |
| Management | *A Theory of Justice* | A classic book. | 5 | 8/10 | 105 |
| LIS | *Understanding Media: The Extensions of Man* | The classic for communication and it is a very important and useful book. | 4 | 1/1 | 63 |
| Psychology | *Introduction to Cognitive Psychology* | The content is novel and profound. There is relevant, valuable | 5 | 1/1 | 71 |
|  |  | information for me. | | | |
| Literature | *Old Tang Records* | Printing and paper is of too bad a quality. | 5 | 9/10 | 346 |

[a] *Macroeconomics* by Rudiger Dornbusch (China Renmin University Press, 1997), *A Theory of Justice* by John Rawls (Harvard University Press, 1971), *Understanding Media: The Extensions of Man* by Marshall McLuhan (McGraw-Hill, 1964), *Introduction to Cognitive Psychology* by Danny Moates (Wadsworth Publishing Company, 1980), *Old Tang Records* by Liu Xun (Zhonghua Book Company, 1975).

The star ratings, review contents, and review helpfulness of each book were then extracted, by parsing the html file crawled from Amazon.cn. Table 2 presents the study's sample following the integration of the citation data with that for review content, star rating, and review helpfulness.

**Method**

The primary purpose of this study is to specify how and whether it is feasible to measure the impact of academic books through the multi-granularity mining of online book reviews. For each book review, review content and a star rating are generated by the review holder. Meanwhile, review helpfulness is generated by review evaluators. Therefore, this study examines, first, the book impact measure and correlation analysis of information from review holders only, and, second, the book impact measure and correlation analysis of information from both the review holder and review evaluators. Fig. 2 summarizes the overall framework of the book impact measure and correlation analysis.

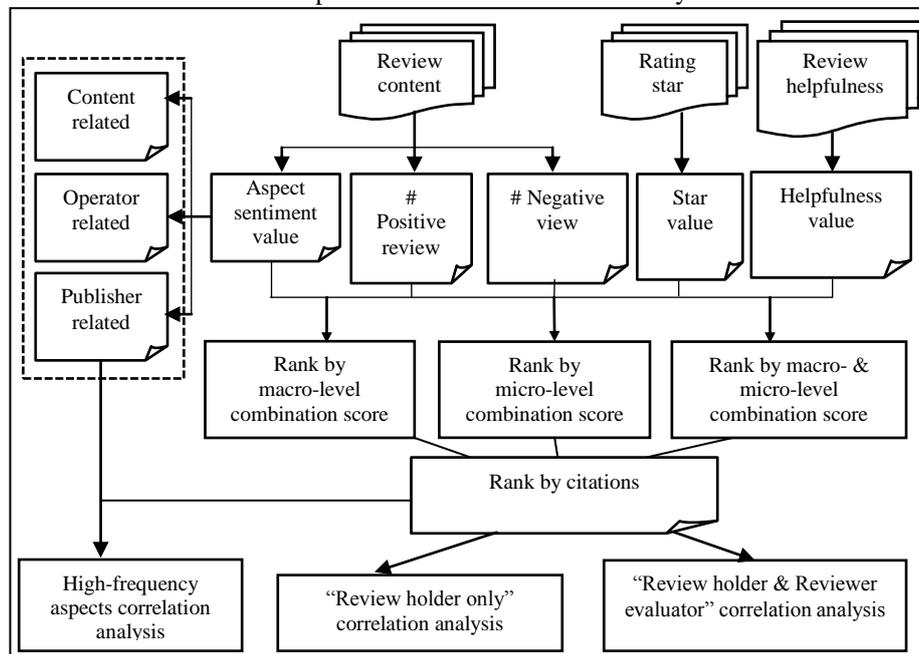

**Fig. 2.** Framework of book impact measure and correlation analysis

The method incorporates three steps, as follows (and discussed in detail in the next section):

(1) *Factor combination.* Different combinations of the factors can reflect different measure results. Star rating, review content, and review helpfulness are the main parts of the Amazon.cn book reviews, and all three are listed separately as well as combined for analysis.

(2) *Factor calculation.* Ratings range from one to five stars on Amazon.cn. Multi-granularity sentiment analysis of review content can be classified into "macro" and "micro" levels. Specifically, macro-level sentiment analysis is conducted to calculate the number of positive and negative reviews, while micro-level sentiment analysis is conducted to extract high-frequency aspects and calculate aspects' sentiment values. Aspects are usually the nouns in the review content, and reflect particular attributes of a book. For example, in the sentence "The content of this book is amazing," the aspect is "content." Review helpfulness is calculated by the number of review evaluations. For example, in "359 of 365 people think it is helpful," the value of helpfulness in the review is 359/365. Finally, the entropy method (Hongzhan et al. 2009) is used to integrate the values to obtain final book impact scores.

(3) *Multi-level correlation analysis.* In order to corroborate the validity of multi-granularity online review mining, correlation analysis between book impact scores and book citations is investigated. In addition, in order to discover which aspects in a review affect citations most, correlation analysis are conducted. Moreover, in order to avoid the effect of the single aspect, the aspects are divided into three categories—content related, publisher related, and operator related—and then the correlation analysis is conducted.

**Impact measure and correlation analysis**
**Factor combination**
Factors considered in this study were the "number of positive reviews," the "number of negative reviews," the "aspect sentiment value," the "star value," and the "helpfulness value." Also, investigations of three levels of combinations were separately conducted for the "Review holder" and the "Review holder & Reviewer evaluator" parts, incorporating a macro-level combination, a micro-level combination, and a macro- and micro-level combination, as shown in Table 3.

**Table 3.** Factor combination

| Factor | Review holder | | | Review holder & Reviewer evaluator | | |
|---|---|---|---|---|---|---|
| | Macro | Micro | Macro & Micro | Macro | Micro | Macro & Micro |
| # Positive reviews | √ | | √ | √ | | √ |
| # Negative reviews | √ | | √ | √ | | √ |
| Aspect sentiment value | | √ | √ | | √ | √ |
| Star value | √ | √ | √ | √ | √ | √ |
| Helpfulness value | | | | √ | √ | √ |

**Factor calculation**
As listed in Table 3, five factors were calculated for the "Review holder" and "Review holder & Reviewer evaluator" parts, respectively (the difference between the two parts being whether or not review helpfulness was taken into consideration), and the computation methods used for each factor is specified in this section.

*Calculating the number of positive reviews and negative reviews.* Macro-level sentiment analysis was used to calculate the numbers of positive and negative reviews. First, word segmentation[3] was applied to the reviews. Then, the "term frequency–inverse document frequency" (TF–IDF) method (Salton and McGill 1983) was applied to the segmented words in order to select feature words, with the aim of improving classification performance. This was calculated using equation (1):

$$\text{TF} - \text{IDF} = \frac{word\_term}{\#word} * log \frac{\#doc}{\#doc\_term} \qquad (1)$$

where TF is "term frequency," and refers to the number of times a given word appears in the document; $word\_term$ is the number of times a word term appears; $\#word$ is the number of words in the document; IDF is "inverse document frequency," a measure of the general importance of words; $\#doc$ denotes the number of documents; and $\#doc\_term$ stands for the number of documents containing the word term.

Finally, a "support vector network" (Cortes and Vapnik 1995) was used to conduct sentiment classification. The sentiment polarity (positive or negative) of each review was identified based on the classification results, and the numbers of positive reviews and negative reviews were then calculated.

*Calculating aspect sentiment values.* A three-step calculation was used to determine the aspect sentiment values, incorporating: (1) aspect extraction, (2) aspect sentiment classification, and (3) aspect sentiment values calculation. In the first step, aspect extraction was conducted in three stages: (a) Chinese word segmentation of the reviews, (b) part-of-speech tagging and the selection of nouns as candidate aspects (as shown in Fig. 3, and (c) TF value calculations of each candidate aspect, and then selecting the top 10 high-frequency aspects.

In the second step, an aspect sentiment classification of each review was computed using equation (2) (Ding et al. 2008) with a sentiment lexicon:[4]

$$sp_{ij} = \begin{cases} +1, \sum_{k=1}^{n} \frac{V_{w_k}}{dis(w_k, s_i)} > 0 \\ -1, \sum_{k=1}^{n} \frac{V_{w_k}}{dis(w_k, s_i)} < 0 \end{cases}, \text{where } i = 1, 2, \ldots, 10, j = 1, 2, \ldots, m \qquad (2)$$

---
[3] Using Jieba (https://github.com/fxsjy/jieba) for the Chinese text segmentation.
[4] Using SentiWordNet (http://sentiwordnet.isti.cnr.it/).

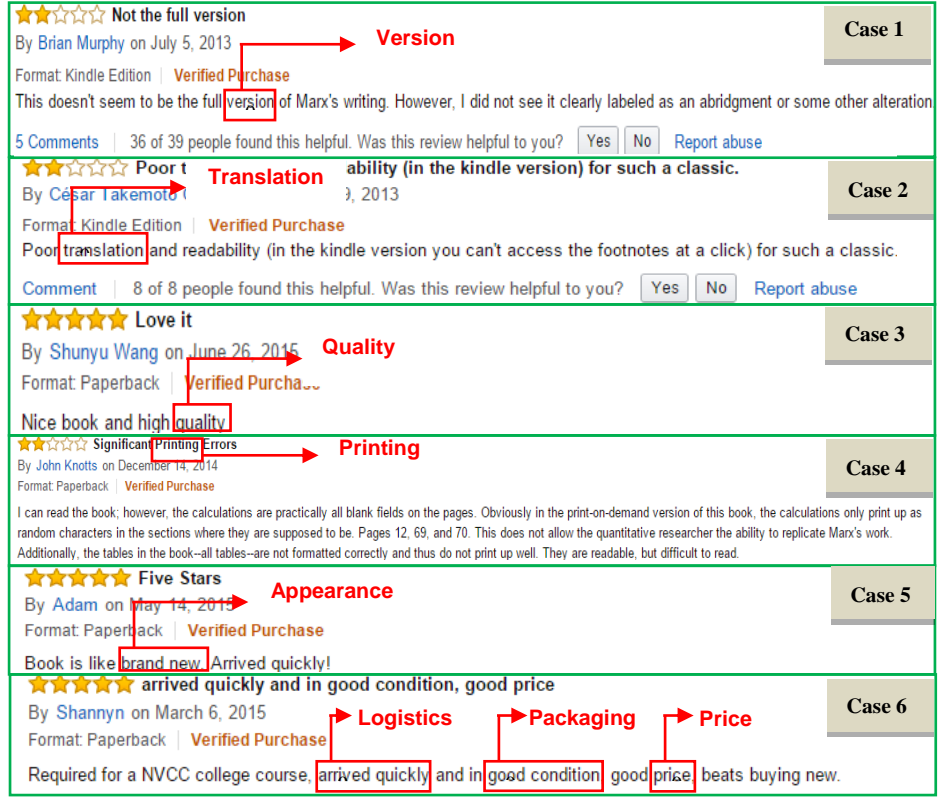

**Fig. 3.** Examples of reviews with high-frequency aspects (the first, second, and third are reviews of Das Kapital, and the fourth, fifth, and sixth are reviews of A Theory of Justice)

where, for the sentiment polarity $sp_{ij}$ of aspect $s_i$ in review $r_j$, $n$ is the number of sentiment words in review $r_j$; $V_{w_k}$ denotes the sentiment value of sentiment word $w_k$, and is equal to +1 if $w_k$ is a positive word, otherwise it is equal to −1; $dis(w_k, s_i)$ denotes the distance between aspect $s_i$ and sentiment word $w_k$; $i$ denotes the number of aspects; and $m$ is the number of reviews. So, for example, in "The content of this book is interesting, but the price is expensive," *interesting* is a positive sentiment word, and $V_{w_k}$ is equal to +1; *expensive* is a negative sentiment word, and $V_{w_k}$ is equal to −1; for the aspect *content*, $\sum_{k=1}^{n} \frac{V_{w_k}}{dis(w_k,s_i)} = \frac{+1}{5} + \frac{-1}{10} = \frac{1}{10} > 0$, $sp_{ij} = 1$; and, for the aspect *price*, $\sum_{k=1}^{n} \frac{V_{w_k}}{dis(w_k,s_i)} = \frac{+1}{3} + \frac{-1}{2} = \frac{-1}{6} < 0$, $sp_{ij} = -1$. Thus, the content in this review was found to be positive, while reference to the price was deemed to be negative.

In the third step, aspect sentiment value was calculated for "Review holder" and "Review holder & Reviewer evaluator," according to whether or not the review's helpfulness was taken into consideration. Equation (3) calculates aspect sentiment value in the "Review holder" part without considering the review's helpfulness:

$$VAB_{it} = \sum_{j=1}^{N} sp_{ij} / \sum_{j=1}^{N} |sp_{ij}|, \text{where } i = 1, 2, \ldots, 10, t = 1, 2, \ldots M \quad (3)$$

where, for aspect sentiment values $VAB_{it}$ of aspect $s_i$ about book $b_t$, $N$ is the number of reviews with aspect $s_i$ about book $b_t$; $i$ denotes the number of aspects; and $M$ is the number of books in each discipline.

In turn, aspect sentiment value in "Review holder & Reviewer evaluator" including review helpfulness was computed by equation (4):

$$VAB'_{it} = \sum_{j=1}^{N} (sp_{ij} * h_j) / \sum_{j=1}^{N} |sp_{ij}|, \text{where } i = 1, 2, \ldots, 10, t = 1, 2, \ldots M \quad (4)$$

where, for aspect sentiment values $VAB_{it}'$ of aspect $s_i$ about book $b_t$, $N$ is the number of reviews with aspect $s_i$ about book $b_t$; $i$ denotes the number of aspects; $M$ is the number of books in each discipline; and $h_j$ is the review helpfulness score of review $r_j$.

*Calculating star value.* Equation (5), without considering review helpfulness, gives the star value calculation for the "Review holder" part as:

$$VSB_{jt} = \sum_{j=1}^{N} star_j / N, \text{ where } t = 1, 2, \ldots M \qquad (5)$$

where, for star values $VSB_{jt}$ of review $r_j$ about book $b_t$, $star_j$ is the star score of review $r_j$, ranging from 1 to 5; $N$ denotes the number of reviews about book $b_t$; and $M$ is the number of books of each discipline.

In turn, star value calculation in "Review holder & Reviewer evaluator," including review helpfulness, was computed by equation (6):

$$VSB'_{jt} = \sum_{j=1}^{N} (star_j * h_j)/N, \text{ where } t = 1, 2, \ldots M \qquad (6)$$

where, for star values $VSB_{jt}'$ of review $r_j$ about book $b_t$, $star_j$ is the star score of review $r_j$, ranging from 1 to 5; $N$ denotes the number of reviews about book $b_t$; $M$ means the number of books of each discipline; and $h_j$ is the helpfulness score of review $r_j$.

*Calculating total score of a book via its reviews.* After the five factors were calculated for the "Review holder" and the "Review holder & Reviewer evaluator" parts, the entropy method was used to calculate the factor weight and total book impact scores (Hongzhan et al. 2009), as shown in Table 4 (equations 7–10).

**Table 4.** Calculating the total book impact scores using the entropy method

| | |
|---|---|
| Step 1: Normalization | $p_{ij} = \frac{v_{ij}}{\sum_{i=1}^{N} v_{ij}}$, where $i = 1, 2, \ldots, N, j = 1, 2, \ldots, m$ |
| Step 2: Factor entropy | $e_j = -\frac{1}{\ln(n)} \sum_{i=1}^{N} p_{ij} \ln(p_{ij})$ |
| Step 3: Factor weight | $w_j = 1 - e_j/m - \sum_{j=1}^{m} e_j$ |
| Step 4: Total book impact scores | $SB_i = \sum_{j=1}^{m} p_{ij} * w_j$, where $i = 1, 2, \ldots, N, j = 1, 2, \ldots, m$ |

Notes: $v_{ij}$ is the value of book $b_i$ in factor $f_j$, $N$ is the number of books, $m$ is the number of factors, $e_j$ is the entropy of factor $f_j$, and $w_j$ is the weighting of factor $f_j$.

**Multi-level correlation analysis**

Two levels of correlations were analyzed in this study. We first conducted a correlation analysis of the book impact scores obtained using our method versus those determined through book citations, trying to find the most correlated combination, and, in turn, prove the reliability of our method. Then, correlation analysis was conducted between the aspect sentiment values of each category and citations of the books, with the aim of discovering which aspect categories most affect book impact.

**Results**

Book rankings as computed by our proposed method are presented in the next section, followed by the results of the correlation analysis of the "Review holder only" and "Review holder & Review evaluator" parts. Finally, the results of the correlation analysis between book citations and sentiment values of high-frequency aspects are presented.

**Book rankings computed by multi-granularity online review mining**

Book rankings obtained using our method have two main parts—"review holders only" and "review holder & review evaluator"—each containing three levels of rankings: macro, micro, and macro and micro. Table 5 lists, as an example, the top 10 economics books, and it can be seen that while, overall, the top 10 books in each level are almost the same, their orders are different. The micro-level combination in "Review holder only" has the highest average (0.7824) and lowest variance (0.0123), suggesting that book impact scores calculated in this combination are generally higher. Scores for *Economics* (2004 edition) and *Economics* (1999 edition) are similar in all six combinations, which seems to indicate that different versions of the same book may get similar public evaluations.

Table 5. Top 10 books of economics discipline in terms of different combinations

| Rank | Book title (citations) | Review holder only | | | Review holder & Review evaluator | | | |
|---|---|---|---|---|---|---|---|---|
| | | Macro level | Micro level | Macro & Micro | Macro level | Micro level | Macro & Micro | |
| 1 | Das Kapital (2261) | Economics (2004 ed.) (0.9999) | Accounting Standards for Business Enterprises (0.9976) | Economics (2004) (0.9898) | Accounting Standards for Business Enterprises (0.9999) | Accounting Standards for Business Enterprises (0.9999) | Economics (2004 ed.) (0.9588) | |
| 2 | Selected Works of Deng Xiaoping (1629) | Economics (1999 ed.) (0.9933) | Das Kapital (0.9436) | Economics (1999) (0.9878) | Das Kapital (0.7983) | Das Kapital (0.7983) | Economics (1999 ed.) (0.9582) | |
| 3 | Competitive Edge (776) | Das Kapital (0.7009) | Methods of Analysis and Modeling on Econometrics (0.8874) | Das Kapital (0.7011) | Management (0.7137) | Management (0.7137) | Das Kapital (0.7019) | |
| 4 | An Inquiry into the Nature and Causes of the Wealth of Nations (683) | Management (0.6714) | Selected Works of Deng Xiaoping (0.7830) | Management (0.6715) | China Economy (0.6949) | China Economy (0.6949) | Management (0.6719) | |
| 5 | Economic Game Theory (240) | A Theory of Justice (0.4459) | Management (0.7709) | A Theory of Justice (0.4461) | Methods of Analysis and Modeling on Econometrics (0.6918) | Methods of Analysis and Modeling on Econometrics (0.6918) | A Theory of Justice (0.4482) | |
| 6 | Supply Chain Management (219) | Macroeconomics (0.3700) | The Global Perspective of Macroeconomics (0.7419) | Macroeconomics (0.3702) | Macroeconomics (0.6631) | Macroeconomics (0.6631) | Macroeconomics (0.3742) | |
| 7 | Economics (1999 ed.) (190) | Microeconomics (0.3498) | Economics (2004) (0.6943) | Microeconomics (0.3499) | A Theory of Justice (0.6262) | A Theory of Justice (0.6262) | Microeconomics (0.3495) | |
| 8 | Economic analysis of property rights (175) | Economic Game Theory (0.3375) | Economics (1999) (0.6943) | Economic Game Theory (0.3376) | Economics (2004 ed.) (0.6200) | Economics (2004 ed.) (0.6200) | Economic Game Theory (0.3390) | |
| 9 | Macroeconomics (143) | Methods of Analysis and Modeling on Econometrics (0.3137) | Privatization and Public–Private Partnership (0.6903) | Methods of Analysis and Modeling on Econometrics (0.3145) | Economics (1999 ed.) (0.6199) | Economics (1999 ed.) (0.6199) | Methods of Analysis and Modeling on Econometrics (0.3199) | |
| 10 | Institutional Economics: Social Order and Public Policy (127) | Structural Equation Model and its Applications (0.2708) | Econometrics (0.6804) | Structural Equation Model and its Applications (0.2709) | An Inquiry into the Nature and Causes of the Wealth of Nations (0.5681) | An Inquiry into the Nature and Causes of the Wealth of Nations (0.5681) | Structural Equation Model and its Applications (0.2749) | |

In order to test the results obtained using our method, a correlation analysis of book impact scores calculated using it versus those found using traditional book citations was undertaken.

**Correlation analysis—"Review holder only"**

A correlation analysis of rankings between the citations approach and the book impact scores obtained via our method was conducted for the "Review holder only" part across three levels of combination—macro, micro, and macro and micro—as shown in Table 6.

**Table 5.** Correlation analysis of citations versus book scores—"Review holder only"

|  | Economics | Management | LIS | Psychology | Literature |
|---|---|---|---|---|---|
| Macro level | 0.370* | 0.340* | 0.274 | 0.165 | 0.188 |
| Micro level | 0.538** | 0.423** | 0.380* | 0.157 | 0.136 |
| Macro & micro level | 0.383* | 0.401** | 0.416** | 0.377* | 0.197 |

\* Correlation is significant at the 0.05 level (two-tailed)

\*\* Correlation is significant at the 0.01 level (two-tailed)

Overall, the book impact scores obtained using our method in the economics and management subject areas have significant positive Pearson correlations with the citations approach in all three combinations. Book scores for LIS publications have significant positive Pearson correlations with citations in both micro-level and macro- and micro-level combinations. Psychology book scores have a significant positive Pearson correlation with citations only in macro- and micro-level combinations, though, while the scores for literature titles show no significant correlation. Reasons for these findings may have to do with the professional degrees corresponding to the books and the educational background of their readers (Zhou and Zhang 2013). That is, books concerning economics, management, and LIS topics are relatively professional and their readers will often follow a definite, protracted educational advancement, whereas books in the psychology or literature categories are typically more popular in style and accessible to a far broader range of readers, each with a wide variety of attitudes and opinions. For the economics and management books in our study, correlations are higher in the micro-level combinations (0.538 and 0.423) and the macro- and micro-level combinations (0.383 and 0.401) than in the macro-level combinations (0.370 and 0.340). For the LIS and psychology titles, correlations are higher in the macro- and micro-level combinations (0.416 and 0.377) and the micro-level combinations (0.380 and none) than in the macro-level combinations. This finding suggests that aspects are important elements in the assessment of a book's impact.

**Table 6.** Correlation analysis of citations versus factor values—"Review holder only"

|  | Economics | Management | LIS | Psychology | Literature |
|---|---|---|---|---|---|
| Star values | 0.265 | 0.283 | −0.200 | −0.065 | 0.022 |
| # Positive reviews | 0.475** | 0.466** | 0.211 | 0.170 | 0.180 |
| # Negative reviews | 0.352* | 0.340* | 0.269 | 0.150 | 0.159 |
| Aspect sentiment values | 0.528** | 0.425** | 0.044 | 0.141 | 0.136 |

\* Correlation is significant at the 0.05 level (two-tailed)

\*\* Correlation is significant at the 0.01 level (two-tailed)

Next, in order to determine the most influential factor for each discipline in the "Review holder only" part, ranking correlation analysis between book citations and values of four single factors was conducted—incorporating star values, number of positive reviews, number of negative reviews, and aspect sentiment values—as shown in Table 7.

Correlations between citations and the values of the four factors are higher for economics (0.475, 0.352, and 0.528) and management (0.466, 0.340, and 0.425) than for LIS, psychology, and literature publications. The highest correlation between citations and factor values in economics is aspect sentiment values (0.528), followed by the number of positive reviews (0.475), while, in management, the highest correlation is the number of positive reviews (0.466), followed by aspect sentiment values (0.425). Hence, it seems that, in the "Review holder only" part, books with more reviews or higher aspect evaluations tend to be cited more often in some disciplines. In addition, factor combination is shown to be useful in measuring a book's impact.

**Correlation analysis—"Review holder & Review evaluator"**
In addition, a correlation analysis of rankings between the citations approach and the book impact scores obtained via our method was also conducted for the "Review holder & Review evaluator" part across three levels of combination—macro, micro, and macro and micro—as shown in Table 8.

**Table 7.** Correlation analysis of citations versus book scores—"Review holder & Review evaluator"

|  | **Economics** | **Management** | **LIS** | **Psychology** | **Literature** |
|---|---|---|---|---|---|
| Macro level | 0.394* | 0.340* | 0.275 | 0.182 | 0.187 |
| Micro level | 0.394* | 0.361* | 0.491** | 0.120 | 0.125 |
| Macro & micro level | 0.378* | 0.417** | 0.380** | 0.409* | 0.240* |

\* Correlation is significant at the 0.05 level (two-tailed)

\*\* Correlation is significant at the 0.01 level (two-tailed)

Overall, book impact scores via our method in macro- and micro-level combinations have significant positive Pearson correlations with citations in all disciplines, although the scores for literature show a less significant correlation. Compared with the results in Table 6, the correlation values shown in Table 8 are an improvement. This signifies that review helpfulness is useful in the correlation results, as well as affirming that online reviews can be used to assess a book's impact.

In terms of differences between the academic disciplines, the correlations between citations and book scores are distinct. In economics, the highest correlations between citations and the three combination levels are seen for the macro-level and micro-level combinations, while, in management, the highest correlation is for macro- and micro-level combinations, followed by micro-level combinations. In LIS, the highest correlations between citations and the three combination levels are seen for micro-level combinations, followed by macro- and micro-level combinations. As for the psychology and literature disciplines, only the macro- and micro-level combinations show a significant correlation. Only the macro- and micro-level combination has a significant correlation with citations in all the disciplines, from which finding it might be inferred that this combination will assess the academic impact of books more accurately.

Next, in order to ascertain the most influential factor for each discipline in the "Review holder & Review evaluator" part, ranking correlation analysis between book citations and the values of four single factors was conducted, as shown in Table 9.

**Table 8.** Correlation analysis of citations versus factor values—"Review holder & Review evaluator"

|  | **Economics** | **Management** | **LIS** | **Psychology** | **Literature** |
|---|---|---|---|---|---|
| Star values | 0.389* | 0.415** | −0.410* | 0.034 | 0.082 |
| # Positive reviews | −0.265 | −0.283 | 0.157 | 0.105 | 0.184 |
| # Negative reviews | −0.475** | −0.466* | 0.269 | 0.150 | 0.161 |
| Aspect sentiment values | −0.352* | −0.340* | −0.099 | 0.116 | 0.187 |

* Correlation is significant at the 0.05 level (two-tailed)

** Correlation is significant at the 0.01 level (two-tailed)

Correlations between citations and the values of the four factors are higher in the economics, management, and LIS subject areas than in psychology and literature. The highest correlation between citations and factor values in economics and management books are the number of negative reviews, followed by star values. For LIS publications, only star values have a significant correlation with citations. Hence, it seems that, in the "Review holder & Review evaluator" part, books with less negative reviews or a lower aspect evaluation tend to be highly cited, while books with higher star evaluations may have different impacts in different disciplines. The results for the psychology and literature titles also indicate that it is necessary to combine the factors, incorporating star values, the number of positive reviews, the number of negative reviews, and aspect sentiment values.

**Correlation analysis—citations versus sentiment values of high-frequency aspects**

The results of our correlation analysis of citations versus the sentiment values of high-frequency aspects are shown in Table 10. The aspect sentiment values were calculated using equation (4), as detailed above.

**Table 9.** Correlation analysis of citations versus aspect sentiment values

| | Economics | Management | LIS | Psychology | Literature |
|---|---|---|---|---|---|
| Quality | −0.112 | 0.233 | −0.104 | 0.133 | 0.150 |
| Content | −0.112 | 0.286 | −0.224 | 0.218 | 0.024 |
| Version | 0.054 | −0.072 | −0.103 | −0.014 | 0.054 |
| Printing | −0.479** | 0.186 | −0.046 | 0.017 | 0.045 |
| Translation | −0.438** | 0.224 | −0.094 | 0.073 | −0.142 |
| Paper | −0.045 | 0.173 | −0.096 | 0.115 | 0.022 |
| Packaging | −0.252 | 0.262 | 0.226 | 0.310 | 0.135 |
| Logistics | −0.166 | 0.263 | −0.167 | −0.024 | 0.158 |
| Price | −0.220 | 0.361* | 0.235 | 0.054 | 0.231* |
| Appearance | −0.093 | 0.361* | −0.020 | −0.087 | 0.122 |

* Correlation is significant at the 0.05 level (two-tailed)

** Correlation is significant at the 0.01 level (two-tailed)

In all academic disciplines, there is no significant correlation between citations and the sentiment values of most aspects, which suggests that grouping and integrating the aspects is necessary. According to the degree of semantic correlation among aspects, we divided the top 10 aspects into three categories—content related, publisher related, and operator related—as shown in Table 11. Note that the aspect "quality" is an abstract noun and always co-concurrent with other frequent aspects, and cannot be distinguished clearly. Thus, it is not summarized into the three categories.

**Table 10.** Categories of high-frequency aspects

| Categories | Aspects | | | | |
|---|---|---|---|---|---|
| Content related | Content | Translation | | | |
| Publisher related | Version | Price | Paper | Printing | Appearance |
| Operator related | Packaging | Logistics | | | |

The results of the correlation analysis of citations and the sentiment values of grouped aspects are presented in Table 12. Overall, aspect sentiment values in economics, management, and LIS show significant Pearson correlations with citations, while aspect sentiment values in psychology and literature show no significant correlation. These results are almost the same as those presented in Tables 8 and 10, supporting the inference that the differences between the disciplines lead to the phenomenon of correlation differences.

**Table 11.** Correlation analysis of citations versus grouped aspect sentiment values

|                   | Economics | Management | LIS      | Psychology | Literature |
|-------------------|-----------|------------|----------|------------|------------|
| Content related   | 0.389*    | 0.437**    | 0.474**  | 0.224      | −0.073     |
| Publisher related | 0.345*    | 0.277      | 0.241    | −0.022     | 0.164      |
| Operator related  | 0.350*    | 0.366*     | 0.159    | 0.183      | 0.167      |

\* Correlation is significant at the 0.05 level (two-tailed)

\*\* Correlation is significant at the 0.01 level (two-tailed)

The highest correlations between citations and grouped aspect sentiment values in the economics, management, and LIS subject areas are content-related aspects (0.389, 0.437, and 0.474), followed by operator-related aspects (0.350, 0.366, and none). This finding may suggest that the content and translations of books are quite important in improving their academic impact, and that choosing a better operator (with good logistics and service) is also imperative.

**Discussion**

This study used online book reviews to assess the impact of academic books. Compared with citation-based evaluation methods, our approach makes use of content information through the application of multi-granularity online review mining. It is thus not simply an analysis of the number of citations, but also an attempt to discover the meaning of each citation by mining content-level information. Thus, we posit that online reviews could prove a useful resource when assessing the academic impact of books.

Compared with evaluation methods based on scholarly reviews, the review corpus for our method is much larger, as online reviews are multiple and updated quickly. With the development of technology concerned with web crawling, the cost of collecting online reviews is reducing. Moreover, online reviews are provided by the public, which reflecting the attitudes and opinions of book users directly. This viewpoint, in contrast to that of traditional impact measurement, leads to a much wider observation of book impacts. Thus, we advance our approach as an additional methodology for use within the toolbox of altmetrics.

For example, for the book *Das Kapital*, our method can not only discover its rank among the books to be assessed, but can also obtain its score (i.e., 0.7019 for the micro- and macro-level combinations in the "Review holder & Review evaluator" part). In addition, we can also calculate the sentiment score of each aspect, such as 0.2694 for content and 0.1071 for price. All of these sentiment scores are useful for improving books, too. For instance, when a book needs to be reprinted, related personnel can become very well informed about which aspects should have more attention paid to them.

Finally, our method is not limited to the assessment of monographs, and can be applied to other forms of books, as well as to other academic entities, such as journals, articles, research papers, reports, and so on.

**Limitations**

Our study is subject to a few limitations. First, its coverage was limited to five academic disciplines (economics, management, LIS, psychology, and literature), which were chosen as academic books in these disciplines usually attract more online reviews than books from other disciplines in the social sciences and humanities fields—hence their selection as a dataset for this, our preliminary study on the topic. Another limitation is that the method presented in this paper can only be used to assess books that have had online reviews prepared in relation to them. Some influential books, such as free EBooks, with no or few reviews cannot be assessed, and may be widely accessed, distributed, and cited without assessable online reviews. Also, the method might be not suitable for measuring new books.

In addition, the credibility of the reviews and our study's single data source may affect the generalizability of the results. Although data pertaining to "helpfulness" is used to reduce the impact of fake reviews, information relating to it is sparse for some books. Moreover, the sample data came from Amazon and thus may lack the diversity of reviews available from other websites. Amazon is an e-commerce website, and reviews featured on it, written by users, are usually quite short and may focus on, for example, the packaging of books, while other websites may feature different characteristic. For example, users of Douban.com might pay more attention to books' contents and prefer to give longer reviews. How best to integrate reviews from different websites is a challenging question for future research.

Finally, the technologies of review mining need to be improved, including the systems for the identification of review sentiment polarities, aspect extraction, and the calculation of aspect sentiment values. In this study, we used a supervised method to classify review sentiments, based on the quality of a tagged corpus; if the scale of the data were to surge, the performance of sentiment analysis might diminish.

**Future studies**

For future studies, following the present work, we propose a few interesting directions. First, scholarly book reviews could be combined with social media book reviews to get more comprehensive assessment results. Expert book reviews are usually provided by professional scholars, who primarily evaluate the contents of the books and assess their academic values; for example, Dugan (2008) commented on the contents of each part of the book *Management Basics for Information Professionals* first, and then evaluated its advantages and disadvantages. By contrast, online review information comes from the wider public, who reflect the different features of the books. When users give online reviews, they may not only consider the contents of a book, but also focus on other aspects, such as printing, packaging, etc. Online reviews of *Management Basics for Information Professionals* from Amazon,[5] in comparison with Dugan (2008), not only reviewed the *content*, but also evaluated the *logistics*, *appearance*, etc. In a further study, scholar book reviews could be collected from websites, such as that for *Choice* magazine (Kousha and Thelwall 2015), and reviews integrated from e-commerce websites and social media, in order to assess books and their impact more comprehensively.

Additional metrics might be considered to validate and strengthen the performance of our method, such as Google Book Search, Web of Science, etc. Also, location information could be taken into consideration to identify user differences across different regions. Sales information for the books, too, could be collected to prove the correlations between citations and the three combination levels discussed above.

Finally, our method is not limited to the assessment of Chinese books, and should be applied to books published in other languages. As diverse language users may possess regional distinctions (Zhou et al.

---

[5] http://www.amazon.com/Management-Basics-Information-Professionals-Edition/product-reiews/1555709095/ref =cm_cr_dp_see_all_btm?ie=UTF8&showViewpoints=1&sortBy=recent (accessed Feb 02, 2016).

2016), difference analysis among different language reviews is also a topic that might usefully be explored further. Moreover, due to the development of the natural language process, our method will become more accurate and operable.

**Conclusions**

This study introduced a framework for measuring book impact according to online reviews and their content. We assessed how online reviews can be used for book evaluations, and found some positive results. In answer to our first research question, online reviews do seem to be valuable for use in the impact assessment of academic books. The multi-granularity mining of such reviews can be applied to identify their sentiment polarities and aspect sentiment values.

Regarding the second research question, content-related aspects have the highest correlation values, followed by operator-related aspects. This suggests that the contents and translations of books are, perhaps unsurprisingly, particularly important in improving the impact of academic books, while choosing a superior operator (with good logistics and service offerings) is also important.

Finally, addressing the third research question, there are clear differences between books published in different academic categories. In economics, management, and LIS disciplines, there were found to be significant positive Pearson correlations between book scores obtained via our method and from citations, while, in the psychology and literature disciplines, there were less or no significant correlations. As theorized above, the reason for this likely lies in the dissimilar readerships (attitudes, opinions, user intentions) and markets for books published in the economics, management, and LIS fields and the more popular titles published in the psychology and literature categories.

As a new and comprehensive perspective for the measurement of book impacts, our method is set to serve as an effective alternative metric with which to assess the qualities of academic publications in the era of Web 2.0 and the development of altmetrics. The theoretical implication of our study lies in the idea that future measurements of altmetrics should incorporate content and sentiment within the web information. The value of motivation analysis is not only revealed in altmetrics, but can also be an important perspective within traditional citation analysis and citation motivation studies. In practice, our method makes enhanced use of current functions and information in social networks and e-commerce sites. Its use could provide interesting solutions at the level of application and inspire more effective communication rules or marketing strategies. As online reviews about authors, journals, and even universities or other institutions can be collected from social media websites, our method has the potential to assess the impacts of these too, although its use has been restricted to measuring book impact in the present study.